\documentclass{cernyrep}
\include{epstopdf}
\usepackage{cite}
\graphicspath{{//cern.ch/dfs/Users/r/rudi/Public/figures/JASpapers/}}
\usepackage{varwidth}
\usepackage{xcolor}

\begin{document}

\title{Introduction to Machine Protection}
\author{R.~Schmidt}
\institute{CERN, Geneva, Switzerland}
\maketitle

\begin{abstract}
Protection of accelerator equipment is as old as accelerator technology and was for many years related to high-power equipment. Examples are the protection of powering equipment from overheating (magnets, power converters, high-current cables), of superconducting magnets from damage after a quench and of klystrons. The protection of equipment from beam accidents is more recent, although there was one paper that discussed beam-induced damage for the SLAC linac (Stanford Linear Accelerator Center) as early as in 1967. It is related to the increasing beam power of high-power proton accelerators, to the emission of synchrotron light by electron--positron accelerators and to the increase of energy stored in the beam. Designing a machine protection system requires an excellent understanding of accelerator physics and operation to anticipate possible failures that could lead to damage. Machine protection includes beam and equipment monitoring, a system to safely stop beam operation (e.g. dumping the beam or stopping the beam at low energy) and an interlock system providing the glue between these systems. The most recent accelerator, LHC, will operate with about $3 \times 10^{14}$ protons per beam, corresponding to an energy stored in each beam of 360~MJ. This energy can cause massive damage to accelerator equipment in case of uncontrolled beam loss, and a single accident damaging vital parts of the accelerator could interrupt operation for years. This lecture will provide an overview of the requirements for protection of accelerator equipment and introduces various protection systems. Examples are mainly from LHC and ESS.
\end{abstract}

\section{Introduction to the school}

This is the first school on beam losses and accelerator protection (in general referred to as machine protection). The school is intended for physicists and engineers who are or may be engaged in the design, construction and/or operation of accelerators with high-power particle or photon beams and/or accelerator subsystems with large stored energy.

We will present the methods and technologies to identify, mitigate, monitor and manage the technical risks associated with the operation of accelerators with high-power beams or subsystems with large stored energy, if failures can result in damage to accelerator systems or interruption of operations. At the completion of the school the participants should be able to understand the physical phenomena that can damage machine subsystems or interrupt operations and to analyse an accelerator facility to produce a register of technical risks and the corresponding risk mitigation and management strategies.

Excellent textbook material and a large number of proceedings from CAS schools are readily available for many topics in accelerator physics and technology (e.g. beam dynamics, synchrotron radiation, superconductivity for accelerators). This is not the case for beam losses and protection of accelerators; there are no books, only a limited amount of lectures in accelerator schools and a few invited contributions to accelerator conferences \cite{Sibley2003,Wenninger2014a,Schmidt2008,Schmidt2013}.

This school is focused on the protection of equipment. Similar strategies are used for the design of systems to protect people and the environment. One lecture during this school will present the challenges for protection of people \cite{Rokni2014}.

An efficient way to learn about machine protection is to consider past accidents and near misses. However, it is not common to present what went wrong during accelerator operation, and it is acknowledged that several examples of past accidents are presented during this school.

The programme of this school follows some questions.
\begin{itemize}
	\item What can go wrong when operating an accelerator?
	\item What are the consequences when something goes wrong?
	\item What mitigation methods can be applied?
	\item What aspects of controls and operation are relevant for machine protection?
\end{itemize}

\section{Introduction to this lecture}

In the first part of this lecture we discuss observations from more than 30 years from various accelerators relevant to the design of machine protection systems. The first time that beam-induced damage was discussed in a paper for the SLAC linac as early as in 1967 \cite{Koontz1967}. Some basic principles for machine protection at today's accelerators are derived. Challenges for machine protection are briefly illustrated with two examples, the CERN Large Hadron Collider (LHC) and the European Spallation Source (ESS). In the second part the performance of accelerators is discussed in relation to hazards and machine protection. A short introduction to machine protection follows. In the last part some accidents at different accelerators are presented.

Accelerators, as all other technical systems, must respect some general principles with respect to safety and protection:

\begin{itemize}
\item Protection of people from different threats (radiation, electrical, oxygen deficiency, etc) has always the highest priority. The main strategy to protect people during accelerator operation is to keep everyone out of defined boundaries around an accelerator when beam is running, ensured by a personnel access system. Usually, protection of people is regulated by the governing bodies;
\item Protection of the environment (e.g. following legal requirements);
\item Protection of accelerator equipment and experiments (the investment).
\end{itemize}

In general, risks come from energy stored in a system (measured in joules) as well as from power when operating the system (measured in watts). Particle accelerators are examples of such systems, since many accelerators operate with a large amount of electrical power (from a few to many MW). The energy and power flow need to be controlled. An uncontrolled release of the energy or an uncontrolled power flow can lead to unwanted consequences such as damage or activation of equipment and loss of time for operation.

Several questions are addressed in this lecture.

\begin{itemize}
	\item What accelerators need protection?
	\item What needs to be protected?
	\item What are the hazards/risks?
	\item What can be the consequences of an accident?
	\item What can be done to prevent accidents?
\end{itemize}

Accelerators can be divided into two classes: accelerators operating with a large amount of stored energy in the particle beam such as synchrotrons and storage rings, and accelerators operating with large beam power such as high-power proton accelerators, free-electron lasers (FELs) and linear colliders.

The energy stored in a particle beam is given by $E_{\rm beam} = N \cdot E_{\rm particle}$, with $N$ the number of particles stored in the accelerator and $E_{\rm particle}$ the kinetic energy of a particle. The beam power is given by the energy per unit of time, $P_{\rm beam} = N \cdot E_{\rm particle} / t$.

For synchrotrons and storage rings, the energy stored in the beam increased over the years (at CERN from the Intersecting Storage Ring ISR to LHC, with an energy of 362~MJ stored in one beam with nominal LHC parameters).

For linear accelerators and fast-cycling machines, the beam power increased over the years. Not only energy or power are relevant but also the beam size. The smaller the beam, the higher is the energy density in case the beam is deflected into equipment. The damage potential increases with decreasing beam size. For accelerators such as future linear colliders, the emittance is expected to become much smaller (down to a beam size of a nanometre) resulting in very high power density (W mm$^{-2}$) \cite{Ross2000,Ross2000a,Jonker2012}.

In order to get an idea of what this amount of energy means, some examples are given. The energy of a pistol bullet is about 500~J; the energy of 1~kg of TNT is about 4~MJ. The energy of 1~l of fuel is about 36~MJ; to melt 1~kg of steel about 800~kJ is required (the energy to melt 1~kg of copper is similar). An accidental release of an energy above 1 MJ can cause significant damage. Even an accidental release of a small amount of energy (order of some hundreds of joules) can lead to some (limited) damage if the energy is released in sensitive equipment such as radio-frequency (RF) cavities.

Protection must be considered during all phases of operation, with the accelerator operating with or without beam. Not only damage to accelerator components needs to be considered, but also to the experiments \cite{Appleby2010}.

Several systems operating with high power or a large amount of stored energy, such as the RF system, power converters, the magnet system and the cryogenic system, are usually commissioned long before beam operation starts.

\section{Synchrotrons and storage rings}

In a synchrotron the beams are injected at low energy and the energy is increased while ramping the magnetic field. The particles are accelerated by RF fields in RF cavities.

\begin{figure}
  \centering
  \includegraphics[width=.45\linewidth]{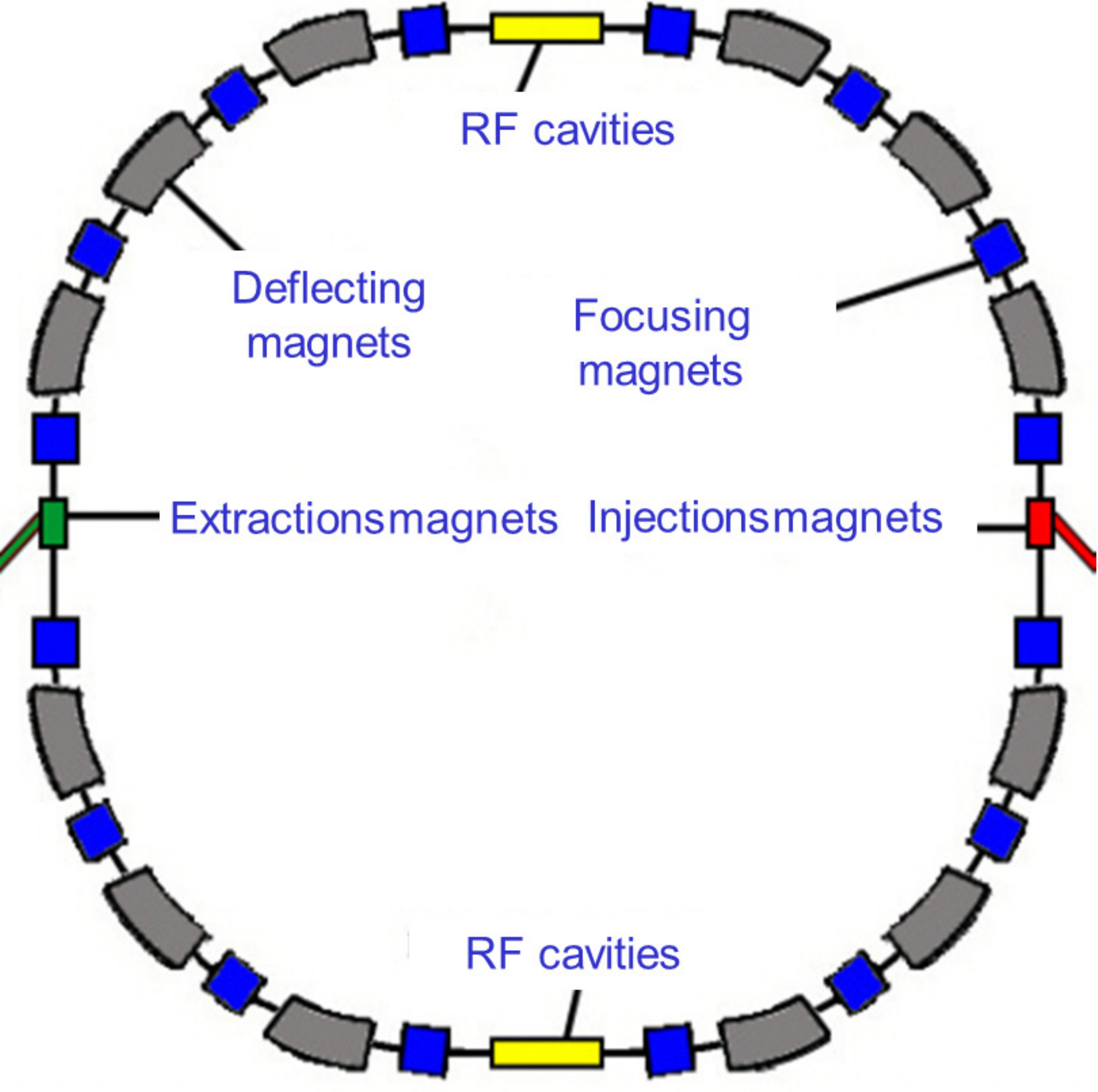}
  \caption{Illustration of a circular accelerator with typical components}
  \label{Components-Circular-Accelerator}
\end{figure}

The main components of a synchrotron are deflecting magnets, magnets to focus the beam and correction magnets (\Fref{Components-Circular-Accelerator}). These magnets are operating with slowly changing fields, on a time-scale of seconds. Pulsed magnets are required for injection and extraction. Power supplies provide the magnet current. Radio-frequency cavities accelerate the beam; the power is provided by components in the RF system (e.g. modulators and klystrons). Other systems are beam instrumentation and the control system. The vacuum system ensures very low pressure for the beam circulating in the vacuum chamber.

Beams are injected at low energy and accelerated to higher energy. Depending on the accelerator, the energy is increased by a factor of 10 to 40. At top energy, the beam is extracted during a plateau to fixed-target experiments or to another accelerator, if the synchrotron is used as pre-accelerator.

For colliders operating with counter-rotating beams, the beams are brought into collision and the length of the plateau is extended to many hours while the beams are colliding (depending on the accelerator, from a few to several tens of hours). For particle physicists, the particle energy and total number of events that are collected are the most important performance parameters. The number of events per unit time is proportional to the luminosity:

\begin{equation}
\frac{N}{\Delta T} = L\,[{\rm cm}^{-2}\,{\rm s}^{-1}] \cdot \sigma\, [{\rm cm}^2],
\end{equation}
with the luminosity $L\,[{\rm cm}^{-2}\,{\rm s}^{-1}]$ and the cross-section $\sigma\,[{\rm cm}^2]$. For head-on collisions the luminosity is given by the number of particles per bunch $N$, the number of bunches per beam $n_{\rm b}$, the revolution frequency $f_{\rm rev}$ and the rms beam sizes at the interaction point $\sigma_x$ and $\sigma_y$:

\begin{equation}
L = \frac{N^2 \cdot f_{\rm rev} \cdot n_{\rm b}}{4 \cdot \pi \cdot \sigma_x \cdot \sigma_y}.
\end{equation}

\begin{table}
\caption{Peak luminosity at different colliders \cite{A.Piwinski.1983,Shiltsev2011}}
\begin{center}
\begin{tabular}{ll}
\hline	
\hline
		{\bf Accelerator}		& {\bf Peak luminosity} 				\\
	\hline		
		PETRA e + e$^-$ (DESY) 				& $2 \times 10^{31}\,[{\rm cm}^{-2}\,{\rm s}^{-1}]$ 	\\
		LEP e + e$^-$ (CERN) 					& $3 \times 10^{31}\, [{\rm cm}^{-2}\,{\rm s}^{-1}]$ \\
		Tevatron  p + p$^-$ (FERMILAB) 		& Some $10^{32}\, [{\rm cm}^{-2}\,{\rm s}^{-1}]$ 	\\
		SuperKEK-B e + e$^-$ (in construction at KEK)  & $10^{36}\, [{\rm cm}^{-2}\,{\rm s}^{-1}]$ 	\\
		FCC e + e$^-$ and pp (study at CERN)  & $5 \times 10^{34}\, [{\rm cm}^{-2}\,{\rm s}^{-1}]$ 	\\	
		LHC (CERN)  					& $1 \times 10^{34}\, [{\rm cm}^{-2}\,{\rm s}^{-1}]$ 	\\	
		HL-LHC (CERN)  					& $2 \times 10^{35}\, [{\rm cm}^{-2}\,{\rm s}^{-1}]$ 	\\	
\hline		
\hline
\label{table:lumitable}			
\end{tabular}
\end{center}

\end{table}

The total number of events, which is what matters for particle physicists, is given by the integrated luminosity, that is, the integral of the luminosity over the time when beams are colliding and experiments take data:

\begin{equation}
N = \sigma \cdot \int{L(t)} \cdot {\rm d}t.
\end{equation}

If the maximum luminosity is limited, either due to accelerator parameters or due to limitations of data taking in the experiments, the efficiency of machine operation plays an essential role for collecting events. The peak luminosity that has been achieved, or that is planned, is given in Table \ref{table:lumitable}.

When a fill is ended, since the luminosity decreased to too low a value, the magnets are ramped down to injection energy and the next cycle starts. The entire process from end collisions to next collisions takes some time (between, say, 30 min and a few hours). Depending on the energy stored in the beams, it might be acceptable that the beams are lost in an uncontrolled way during the down-ramp, or the beams must be extracted to safely deposit the energy in a beam dump block. If the beam is lost during a fill due to a failure a new cycle starts, but the efficiency for data taking is reduced.

\subsection{Luminosity and consequences for machine protection}

The performance of an accelerator is determined by beam parameters and parameters of the hardware systems. For a circular accelerator, energy and luminosity are the most important parameters. The energy stored in the beam as a function of luminosity and accelerator parameters can be approximated by rewriting the luminosity equation:

\begin{equation}
E_{\rm beam} = \frac{\sigma E C}{c} \cdot \sqrt{\frac{4  \pi L}{\delta T}},
\end{equation}

\begin{equation}
E_{\rm beam} = \sigma E \cdot \sqrt{\frac{4 \pi L n_{\rm b}}{f_{\rm rev}}},
\end{equation}
with $\sigma$ the beam size at the interaction point assuming round beams, $E$ the energy, $C$ the circumference, $c$ the speed of light, $L$ the luminosity, $f_{\rm rev}$ the revolution frequency and $\delta T$ the time between two bunches. This is an approximation since it assumes that the machine is filled without any large gaps between bunches. Such gaps, e.g. with a length of some 100~ns or a few $\mu$s, are required for injection and for extraction. As an example, we assume the nominal parameters for LHC: luminosity, $L=10^{34}$ cm$^{-2}$s$^{-1}$; number of bunches per beam,  $n_{\rm b}=2808$; revolution frequency, $f_{\rm rev}=11$ kHz; energy, $E=7$ TeV; and beam size at the interaction point, $\sigma = 16$ $\mu$m. With the approximate equation the energy stored in the beam is $E_{\rm beam}=407$ MJ; the exact calculation yields 362 MJ.

In order to reach high energy, the particles are deflected with superconducting magnets. The Lorentz force on a charged particle is proportional to charge, electric field and the vector product of velocity and magnetic field. An approximation for the energy stored in superconducting magnets is given by the energy of the magnetic field in the vacuum chamber (this is a lower limit; the equation is for a superconducting magnet with two vacuum chambers as in LHC):

\begin{equation}
E_{\rm magnets} = \frac{2 \cdot {\rm Length} \cdot R^2 \cdot \pi \cdot B^2} {\mu_0}.
\end{equation}

For the LHC, the length of the dipole magnet systems is about 20 km, the radius of the vacuum chamber of $R = 28$ mm and the magnetic field of $B = 8.3$ T for an energy of $E = 7$ TeV. This approximation gives an energy stored in the LHC dipole magnets of 4.8~GJ. The exact calculation gives 8.82~GJ ($\mu_0$ is the permeability).

\begin{figure}
  \centering
  \includegraphics[width=.75\linewidth]{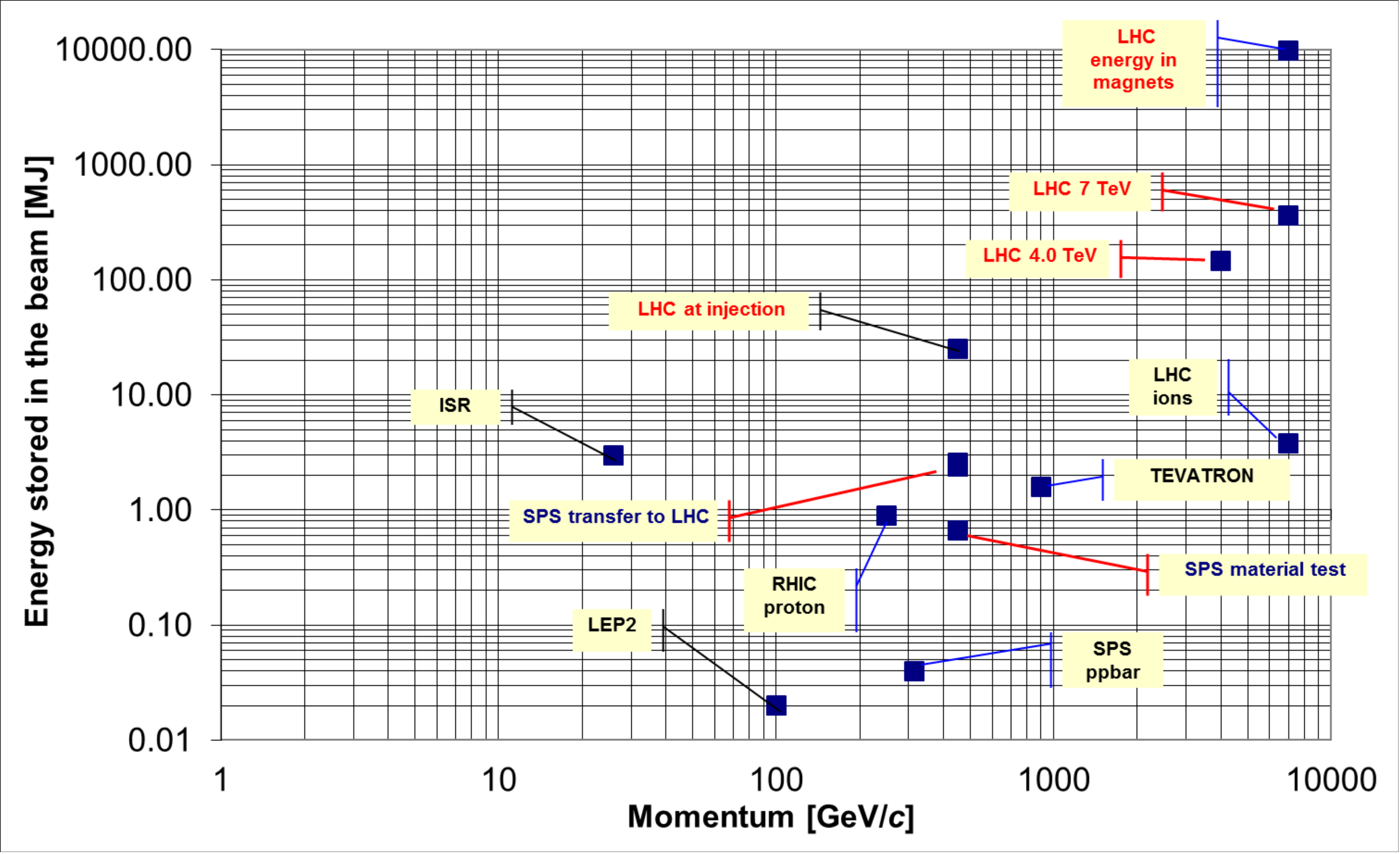}
  \caption{Energy stored in the beams for different accelerators and the energy stored in the LHC magnet system}
  \label{fig-stored-beam-energy}
\end{figure}

The energy stored in the beam for different accelerators as well as the energy stored in the LHC magnet system is shown in \Fref{fig-stored-beam-energy}.

\section{High-power hadron accelerators}

\begin{figure}
\centering
\includegraphics[width=0.7\linewidth]{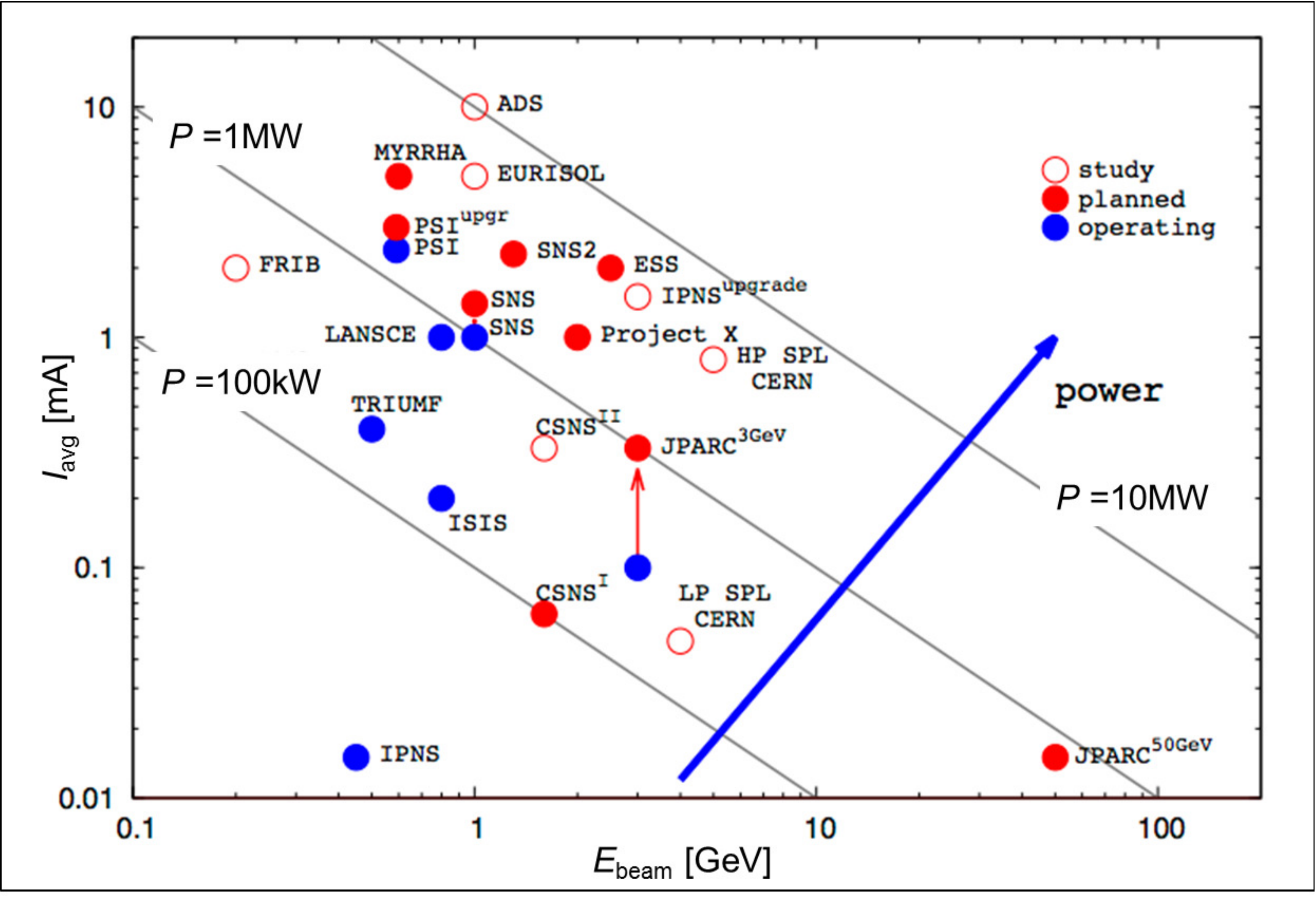}
\caption{Current versus particle momentum for high-power proton accelerators around the world}
\label{High-Power-Accelerators}
\end{figure}

There is a large interest in the exploitation of high-power hadron accelerators. In spallation sources high-intensity proton beams are accelerated and directed to a target. The protons interact with the target material and spallation neutrons are produced. Other accelerators are using high-intensity proton beams for neutrino production. Rare-isotope beams are produced by accelerating ions (e.g. the Facility for Rare Isotope Beams, FRIB, at Michigan University is a folded linac to accelerate ions). Accelerator-driven systems (ADS) are being developed with several projects around the world. A very energetic particle beam is used to stimulate a reaction in a subcritical reactor, which in turn releases enough energy to power the particle accelerator and leaves an energy profit for power generation. Figure \ref{High-Power-Accelerators} shows beam current, beam power and particle momentum for different high-power proton accelerators.

There is a difference between accelerators operating with high-power beams and those with large stored energy. For hadron colliders, the energy stored in the beams can be very high, as has been shown for the LHC. In case of a failure, the energy stored in beam and magnets needs to be safely deposited.

For high-power accelerators, the beam power increases along the accelerating structure proportionally to the particle momentum. The energy stored in the particles present in the accelerator at one moment in time is small. In a case of failure, e.g. causing beam losses in one linac section, the beam must be stopped. It is straightforward to stop the production of particles at the source. After stopping the particle production in the source there are still particles between the source and the location of the beam loss to be considered.

\section{Lessons learned at some accelerators}

In this section, observations related to machine protection for various circular accelerators are presented. These considerations allow us to draw some conclusions that were fundamental for the design of the LHC machine protection systems, as well as for the design of the protection systems for high-power linear accelerators such as ESS.

\begin{itemize}
	\item DESY PETRA I,  e + e$^-$ collider (1978 to 1986) and PETRA III (in operation).
	\item CERN SPS,  proton--antiproton collider (1982 to 1990).
	\item CERN SPS,  proton synchrotron (starting from 1978, still operating).
	\item CERN LEP,  large electron--positron collider (1989 to 2000).
\end{itemize}

\subsection{PETRA, an electron--positron collider at DESY}

PETRA was built as an electron--positron collider (PETRA I). Later PETRA II was used as injector for HERA (Hadron Accelerator Ring Anlage) and now it operates as an advanced synchrotron light source. PETRA has a length of 2304~m and the particles are deflected with normal-conducting magnets.  PETRA~I operation started in 1978 and the particle momentum was up to 21~GeV/$c$. PETRA~I was operating with four bunches per beam, each beam with a current of about 6.5~mA. Frequently the beam was lost, sometimes for unknown reasons, in particular during the energy ramp, which is always very critical since parameters such as betatron tune and chromaticity need to be accurately controlled. When the beams at PETRA~I were lost there was no risk of damage, since the stored beam energy was very small.

One way of stopping the beams in an electron--positron accelerator is to switch off the RF field. The particles are lost gradually due to the energy loss by the emission of synchrotron radiation. The particles are distributed around the circumference and the energy deposition in accelerator equipment is very low.

Equipment protection at PETRA I was important. An example of an incident: for measuring beam polarization by analysing the Compton-scattered photon distribution, a high-power laser beam was sent into the vacuum chamber through a glass window. The glass cracked when the laser passed the window, the vacuum pressure increased and an intervention was required. The window could be sealed and operation continued without major impact.

\begin{figure}
\centering
\includegraphics[width=0.77\linewidth]{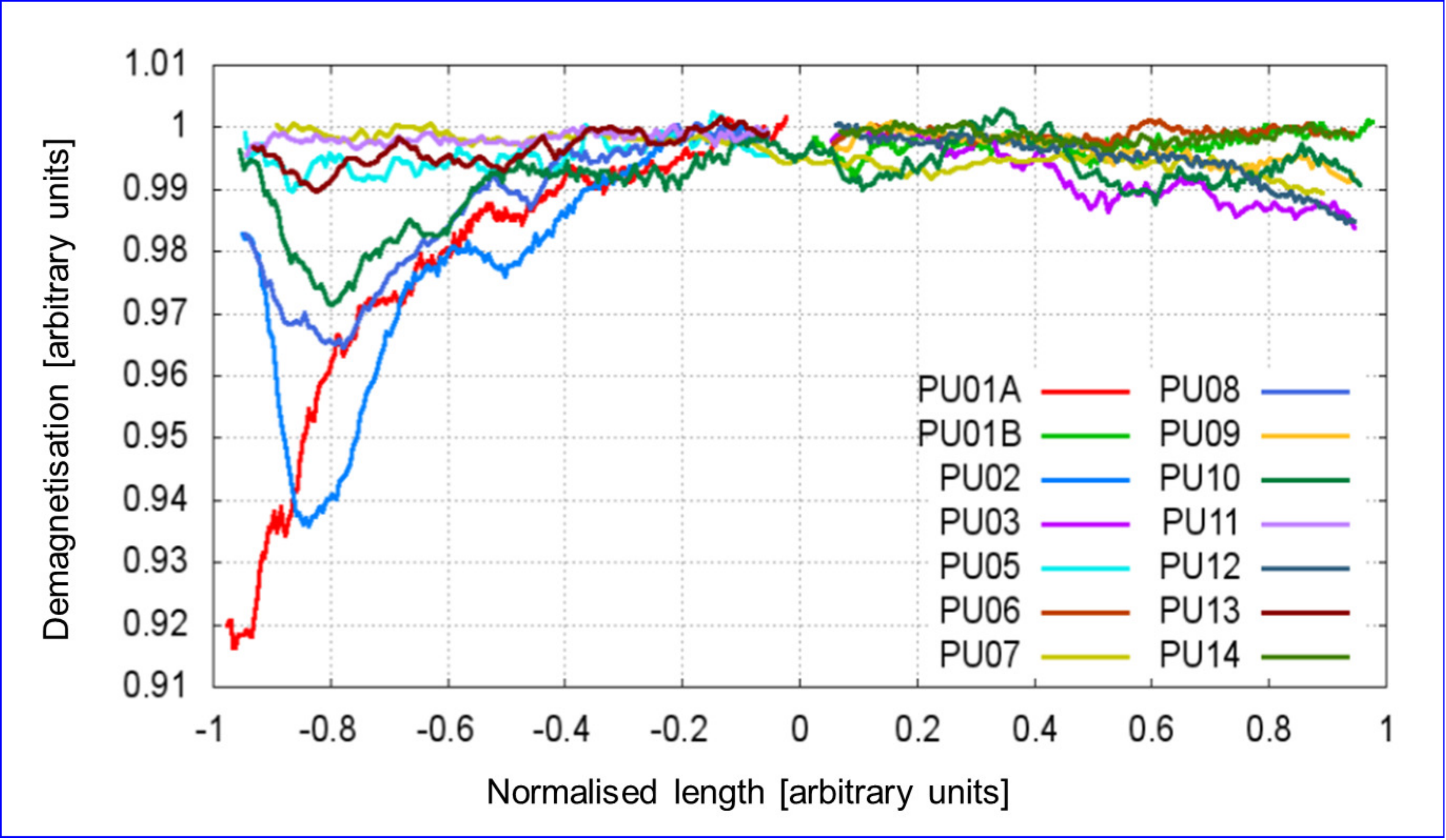}
\caption{Demagnetization of undulators at the PETRA storage ring from \cite{Vagin2014}}
\label{PETRA-damage-undulator}
\end{figure}

PETRA III has been operating since 2008 as a world-class synchrotron light source. Injection is at operating energy and no energy ramp is required. A degradation of the performance of undulator magnets was observed, a de-magnetization that is likely due to beam losses (\Fref{PETRA-damage-undulator}, taken from \cite{Vagin2014}).

\subsection{CERN-SPS synchrotron and proton--antiproton collider}

The SPS was built as a normal-conducting proton synchrotron and operation started in 1978. A proton beam was accelerated to 450 GeV and directed onto a target for fixed-target experiments. Since the beams are circulating for only a few tens of seconds, no ultra-high vacuum was required. The SPS accelerator was transformed into a proton--antiproton collider during 1980 to 1982, and continued operating in this mode until 1990. Antiprotons are very rare and it took a long time to produce them in sufficient quantity in the pre-accelerators to fill the SPS with antiproton beam. If a fill was lost in the SPS, it took many hours to accumulate enough antiprotons for refilling the machine. Therefore, protecting the beam (avoiding an accidental loss of the fill) was of high priority. The energy stored in the beam was not very high, in particular at injection energy; however, on one occasion, injected beam was deflected for a period of about 10~min into the UA2 experiment and led to a degradation of sensitive parts of the experiment (see below).

From 1990 the SPS continued to operate as a synchrotron for fixed-target physics, neutrino production and as injector for LHC. The parameters for the operation as a synchrotron are very different from collider parameters with much higher beam intensity. The beam current in the SPS was constantly being increased over the years. The cycle time was of the order of some seconds to some tens of seconds. If the beam is accidentally lost during one cycle, the efficiency is hardly affected since the next cycle follows a few seconds later. However, beam losses risk to damage components and activate accelerator equipment. Protecting the equipment from beam-induced damage is required and the beams must be extracted into a beam dump in case of a failure. The SPS is operating in different modes with different extraction lines. Beams are extracted with different energies and safe operation requires a complex protection and interlock system. During the operation of the SPS as a synchrotron, magnets and other equipment were damaged several times due to accidental beam losses. In general, it was possible to repair the damage within a short time (< one day), e.g. replacing a magnet since the accelerator is normal conducting. Such repair would be very different for a superconducting machine; the consequence of damaging a superconducting element requires several months of repair. The SPS protection systems have been upgraded following the development of the LHC machine protection; since then there have been no accidents.

\subsection{CERN-LEP electron--positron collider}

LEP (Large Electron Positron collider), operating from 1989 until 2000, was installed in the tunnel that is now used for LHC. The particle energy was up to 104~GeV; the beams were injected at an energy of 20~GeV. Initially, the maximum energy was limited to about 50~GeV and it was sufficient to dump the beams after the end of a fill by switching off the RF field.

Improvement of LEP performance (higher energy and higher intensity) made it necessary to dump LEP beams in a fully controlled way. Fast kicker magnets, close to defocusing QL8 quadrupoles, vertically deflected the bunches; other quadrupoles gave an additional vertical deflection to send the beams into absorbers \cite{E.Carlieretal.1994}.

In June 1996, operation was just about to begin with the upgraded machine when an unexpected problem appeared. Operators were injecting beam, but it was not getting around the whole accelerator. After careful investigation, the cause was found: a pair of Heineken beer bottles wedged into the beam pipe. Beam pipes at LEP were easily accessible and repairs could be made fairly quickly.

An issue for e + e$^-$ beams is the emission of synchrotron radiation by charged particles with increasing energy. The power emitted by one particle circulating  in a dipole field with the bending radius $\rho$, the mass $m$, the charge $e_0$ and the energy $E$  is given by

\begin{eqnarray}
P_s = \frac{e_0^2 \cdot c}{6 \pi \epsilon_0 \cdot (m c)^4} \cdot \frac{E^4}{\rho^2},
\end{eqnarray}
where $c$ is the speed of light and $\epsilon_0$ the vacuum permeability.

The radiation power increases with the energy as the power of four. This became an issue when the energy of LEP was increased from 50~GeV to more than 100~GeV \cite{R.Baileyetal.1998}. Assuming a beam current of 5~mA per beam, the power emitted by both beams at 45~GeV is 770~kW and at 90~GeV it is 12~MW. When the beam passes through a wiggler magnet (a series of magnets designed to periodically laterally deflect the beam), the power density is further increased. Wiggler magnets were used at LEP and are commonly used in synchrotron light sources and free-electron lasers such as the European XFEL at DESY.

When the energy of LEP was increased above 80~GeV, lead stoppers located in front of the aluminium windows of the polarimeter were installed to protect the device when it is not in use. After 30 days running at 80.5~GeV per beam, several of these blocks were found melted and needed to be replaced by tungsten. Other equipment damage was observed: beam instrumentation, electrostatic separators, vacuum equipment and, in particular, wiggler magnets.

Emission of synchrotron radiation is a feature of normal operation. This radiation can damage equipment and needs to be taken into account in the design and during upgrades. This is not limited to LEP, but a potential issue for all synchrotron light sources and XFELs.

At LEP, the beam was frequently lost, e.g. during the energy ramp, without understanding the mechanisms for the loss. No adequate diagnostic system was available to record beam and hardware parameters. A system recording the equipment and beam parameters, including when the beam was lost, would have been very useful (post-mortem system).

\subsection{Some lessons for the design of new accelerators}

From the experience at these accelerators, as well as from many other machines not mentioned here, some lessons can be derived.

\begin{itemize}
\item Protection of equipment is required when there is a significant amount of energy stored in an accelerator system (e.g. superconducting magnets), or if the accelerator operates with high power (e.g. RF systems).
\item Accelerator equipment as well as experiments require protection from uncontrolled beam losses in case of a significant amount of energy stored in the beam, or for high-power beams.
\item For lepton accelerators, the power of synchrotron radiation needs to be considered in order to avoid possible damage.
\item There is the risk of performance degradation for undulators using permanent magnetic material.
\item It is important to understand what happens, e.g. when the beam is lost.
\end{itemize}

\section{Accelerators with machine protection challenges}

\subsection{The CERN-LHC large hadron collider}

The LHC is designed to operate at a momentum of 7~TeV/$c$ with 2808 bunches, each bunch with a nominal intensity of $1.15 \times 10^{11}$ protons. Machine protection is required during all phases of operation, since the LHC is the first accelerator with injected beam far above the threshold for damage. The energy stored in the nominal LHC beam of 362~MJ corresponds to the energy of a 200~m long fast train at 155 km per hour and to the energy stored in 90~kg of TNT. Surprisingly, this is the same as the energy stored in 15~kg of chocolate; it matters most how easily and fast the energy is released. The energy in an accelerator beam can be released in some 10~$\mu$s.

At 7~TeV/$c$, fast beam loss with an intensity of about 5\% of a single `nominal bunch (10$^{11}$ protons)' could damage equipment (e.g.\ superconducting coils). The only component that can stand a loss of the full beam is the beam dump block. All other components would be severely damaged. The LHC beams must \textit{always} be extracted into the beam dump blocks at the end of a fill as well as in case of a failure.

The LHC is a two-ring collider (two different vacuum chambers with opposite magnetic fields for deflecting the particle trajectories) using superconducting magnets. Superconducting magnets store a large amount of energy and need to be protected in case of a quench. Protection of superconducting magnets is part of the design of the magnets, taking into account that most magnets are powered in series \cite{Pfeffer2014}. For the LHC, quench heaters and energy extraction systems are used to prevent magnet damage after a quench. Quench heaters are stainless steel strips installed in the magnet that are heated by discharging the energy stored in capacitors into the strips when a quench is detected. Energy extraction is provided by switching a resistor in series with the magnet chain to extract the energy from the magnets.

Protecting equipment from uncontrolled release of beam after a failure requires very complex systems. Detection of failures is being done by many different monitors providing interlock signals. There are potentially many thousands of such interlock signals from various systems around the accelerator. As an example, in case of a failure in the magnet powering system, e.g. after a quench, the beams must be extracted. Managing of the different interlocks with the objective to build a coherent system is required.

The development of the machine protection system requires expertise in several domains: physicists defining failure scenarios using computer tools, e.g. for particle tracking and particle--material interactions, systems experts working on machine equipment, and physicists or engineers responsible for the operation of such a complex machine.

The machine protection systems traverse the organization. Therefore, at CERN the co-ordination of all activities related to LHC machine protection was assigned to a team in 2000, the Machine Protection Working Group that co-ordinated design, implementation, commissioning and operation of the machine protection system.

\subsection{The European Spallation Source}

\begin{figure}
\centering
\includegraphics[width=1.0\linewidth]{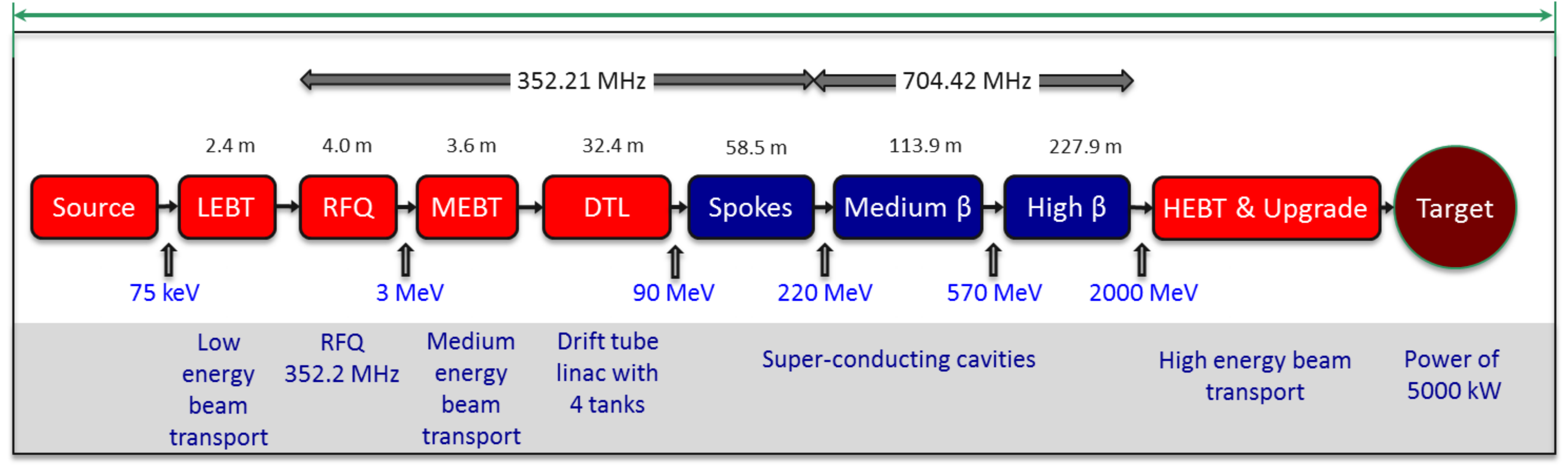}
\caption{Layout of the ESS accelerator: the source, low energy beam transport and RFQ are followed by the medium energy beam transport. The protons are accelerated by a normal-conducting linac, followed by three sections of superconducting cavities. In the high energy beam transport line the protons are transported to the target.}
\label{ESS-layout}
\end{figure}

The European Spallation Source (ESS) being built at Lund, Sweden is designed to accelerate a proton beam with an average power of 5~MW and to direct the protons onto a target. Operation of the ESS will be at a frequency of 14~Hz, with a pulse length of 2.86~ms and a peak power of 125~MW. The layout of the ESS accelerator is shown in \Fref{ESS-layout}.

\begin{figure}
\centering
\includegraphics[width=0.65\linewidth]{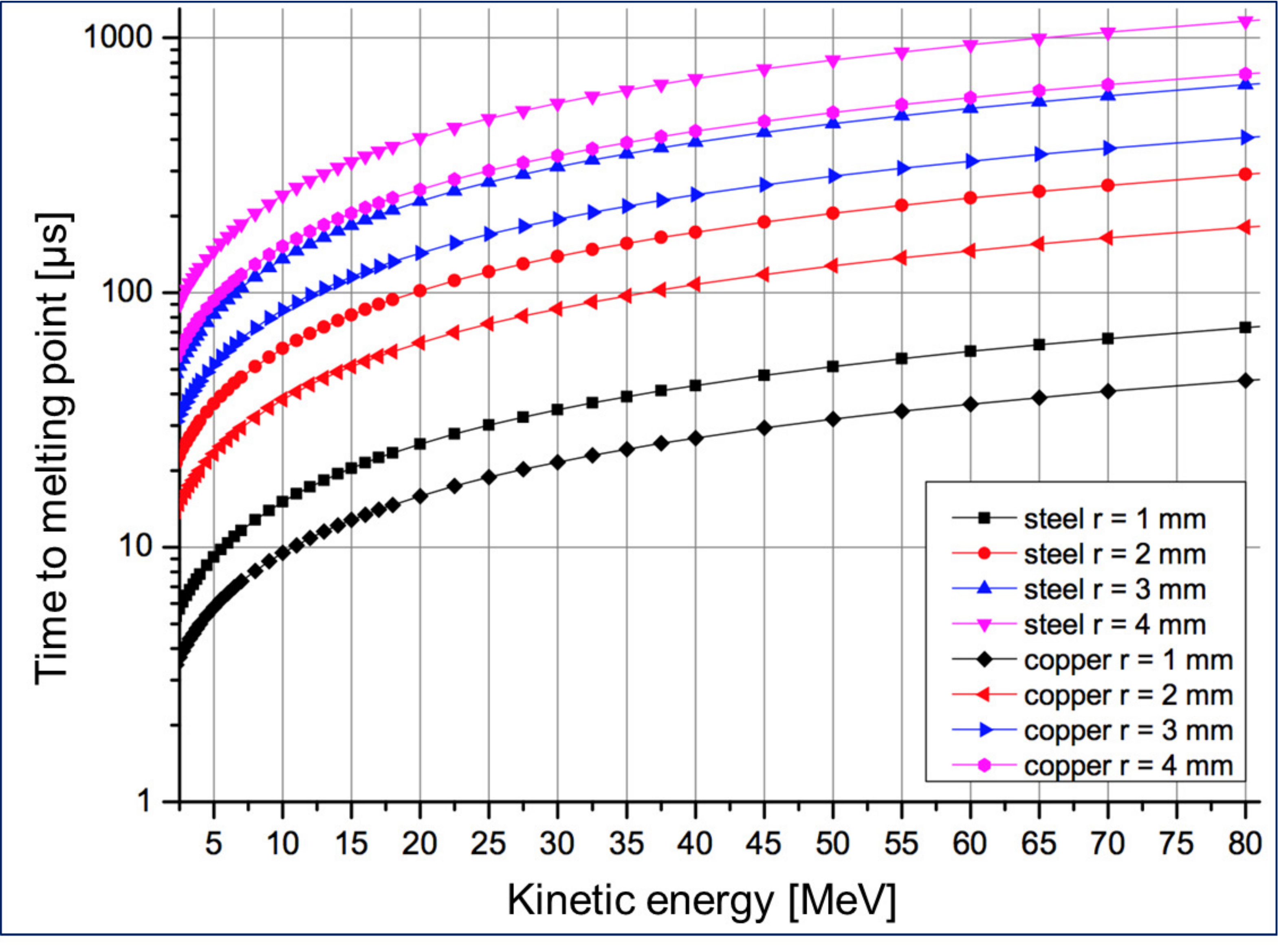}
\caption{Time to melt copper and steel, as a function of proton momentum for different beam sizes \cite{Tchelidze2012}}
\label{ESS-Laliplot}
\end{figure}

\begin{figure}
\centering
\includegraphics[width=0.8\linewidth]{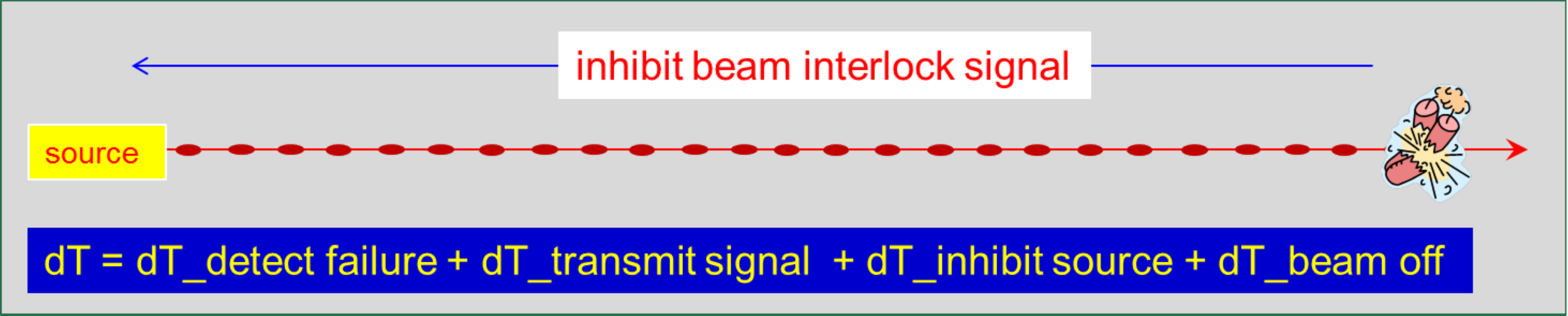}
\caption{Time from detecting a failure until no particles are present in the accelerator: time to detect a failure: dT$_-$detect failure, time to transmit the signal to the source: dT$_-$transmit signal, time to inhibit proton production: dT$_-$inhibit source, time until no protons are in the accelerator: dT$_-$beam off.}
\label{ESS-Interlock-Delay}
\end{figure}

In case of an uncontrolled beam loss during, say, 1~ms at ESS the deposited energy is up to 130~kJ, for 1~s up to 5~MJ. It is required to inhibit the beam after detecting uncontrolled beam loss as fast as possible. There is some delay between detection of a failure (e.g. detection of beam losses by a beam loss monitor) and `beam off'. Figure \ref{ESS-Laliplot} shows the time to melt copper and steel in the case where the proton beam hits a metal surface between 3 and 80~MeV/$c$ \cite{Tchelidze2012}. For example, after the Drift Tube normal-conducting Linac (DTL), the proton energy is 78~MeV/$c$. In case of a beam size of 2~mm radius, melting would start after a beam impact of about 200~$\mu$s. Inhibiting of the beam after a failure is detected should be in about 10\% of this time (\Fref{ESS-Interlock-Delay}, see \cite{Nordt2014}).

\begin{figure}
\centering
\includegraphics[width=0.6\linewidth]{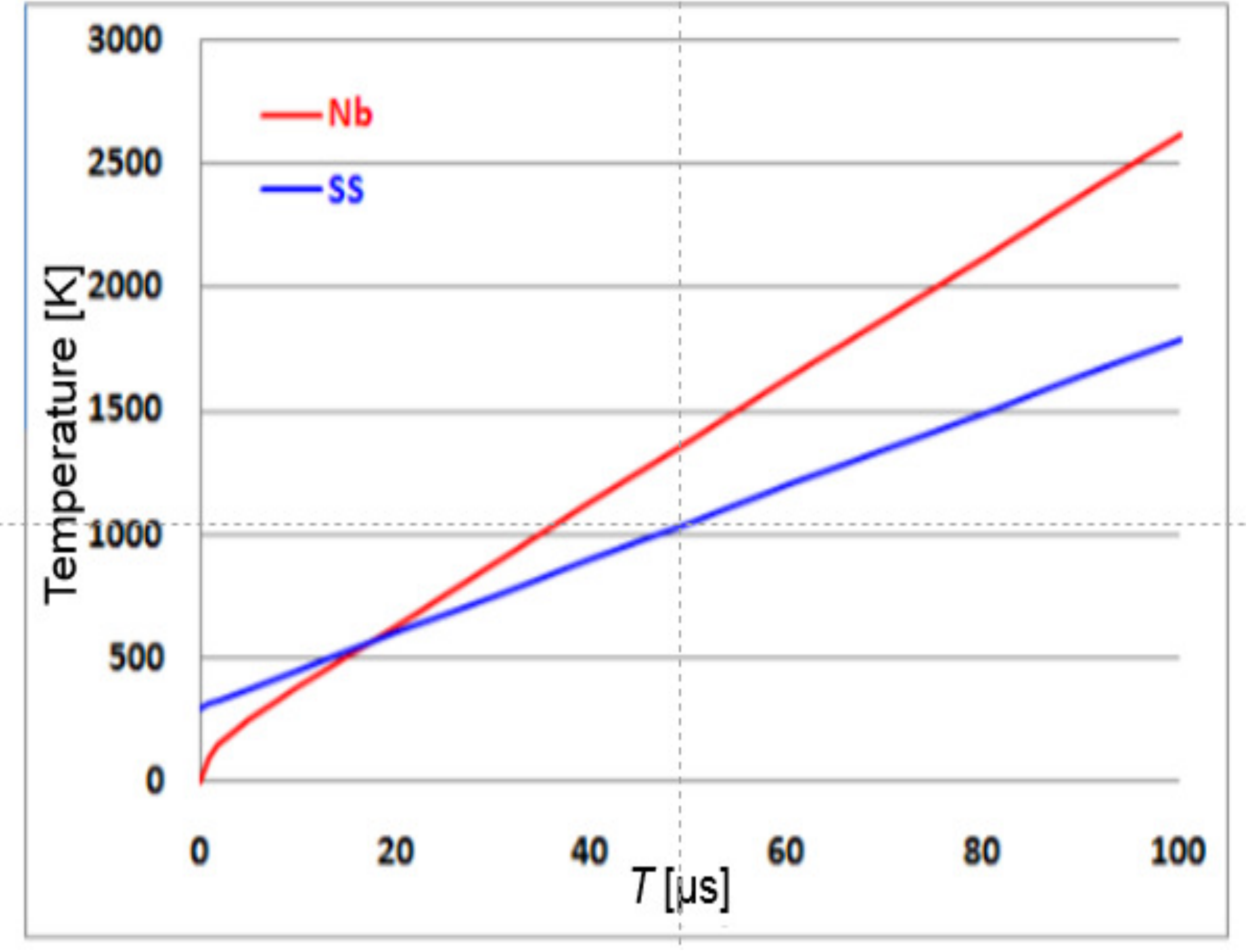}
\caption{Temperature versus time of stainless steel (SS) from 300~K and niobium (Nb) from 2~K, after being hit by a uranium beam, 100~MeV/u, 200~kW and beam rms radius 1 mm \cite{Zhang2011}.}
\label{FRIB-damage}
\end{figure}

As a comparison, a prediction of the damage limit from FRIB beams is shown in \Fref{FRIB-damage}.

\section{The performance of an accelerator and its availability}

For particle colliders, the total number of events is proportional to the integrated luminosity. For many years the most relevant parameter for the operation of a collider was the peak luminosity. The most significant improvements of the integrated luminosity was achieved by increasing the peak luminosity, e.g. at LEP, at the SPS proton--antiproton collider and at the Tevatron \cite{Shiltsev2011}. The accelerators were not as complex as today and availability was not a major concern. With more complex accelerators and limitation of the maximum luminosity this is changing.

\begin{figure}
\centering
\includegraphics[width=0.7\linewidth]{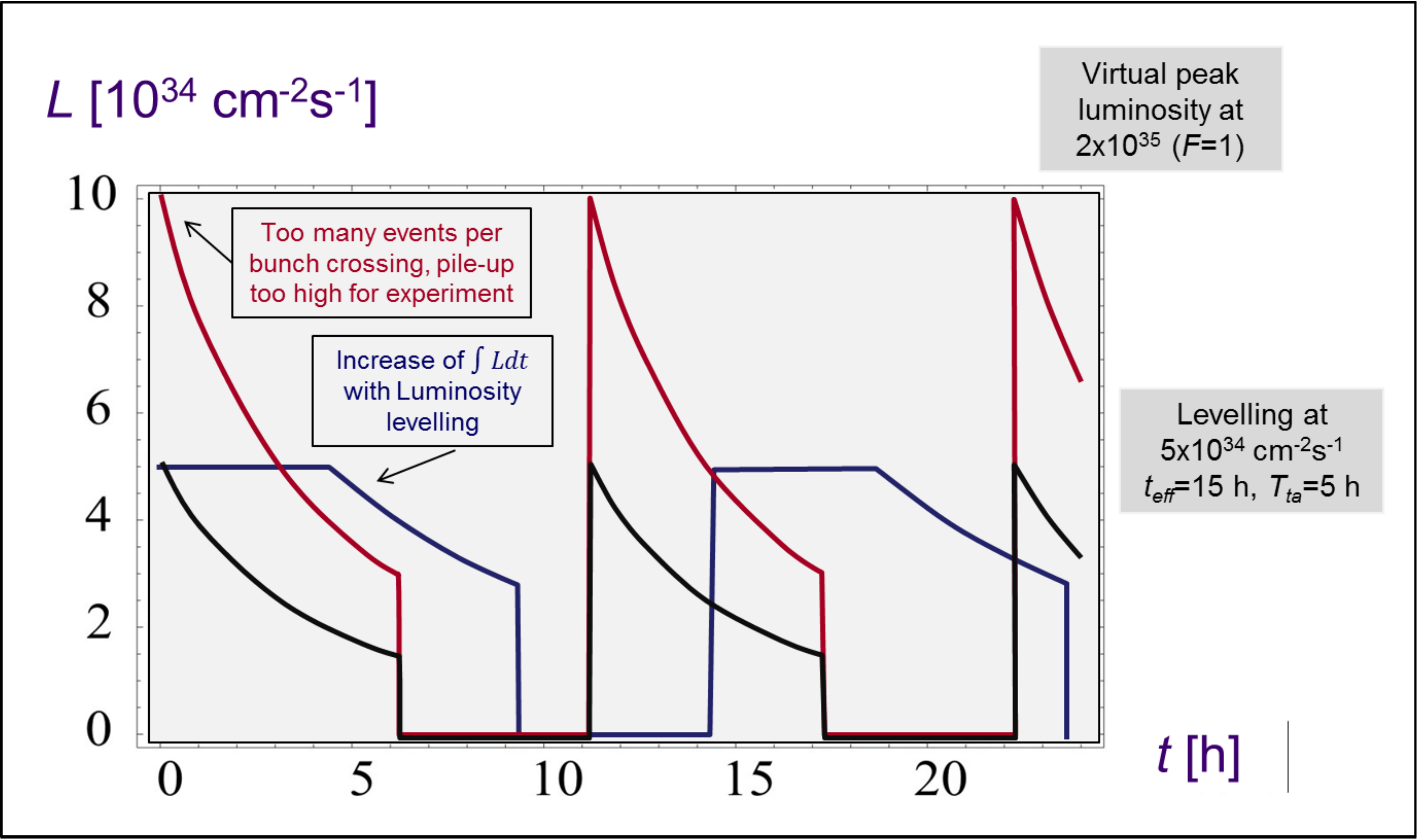}
\caption{Luminosity levelling at HL-LHC}
\label{LHC-lumi-levelling}
\end{figure}

HL-LHC (high-luminosity LHC) is a project aiming at an increase of the integrated luminosity of the LHC per year by a factor of 10, from 20--30 fb$^{-1}$ to more than 300 fb$^{-1}$ per year. In principle, with the HL-LHC beam parameters a luminosity of $2 \times 10^{35}$ [cm$^{-2}$ s$^{-1}$] could be achieved. With a luminosity of $5 \times 10^{34}$ [cm$^{-2}$ s$^{-1}$] the number of events per bunch crossing is of the order of 140 (today it is about~30). For the LHC experiments a further increase of the pile-up beyond, say, 140, is not acceptable. In order to achieve the target integrated luminosity, it is planned to level the luminosity during the fill (\Fref{LHC-lumi-levelling}).

The integrated luminosity depends therefore on the time that the beams are colliding and the experiments take data. The time is directly related to the availability of the accelerator.

For neutron and neutrino sources an important figure of merit is the integrated number of protons on target. Similar quantities can be defined for other accelerators (e.g. synchrotron light sources). For accelerators with many small experiments that take data only for a few days, the availability is important for another reason: if the accelerator is down and not providing beam, users can lose their entire data-taking period.

Damage due to accidental release of energy stored in the beam or in equipment has a major impact on the availability.  Machine protection systems prevent such damage. However, machine protection is not an objective in itself; it is to maximize operational availability by minimizing downtime (quench, repairs, waiting for cool-down to access equipment) as well as to avoid expensive repair of equipment and irreparable damage. Since all technical systems cause some downtime, machine protection systems will also contribute to downtime. For an accelerator with limited risk, the presence of protection systems might reduce the overall availability. If the risk is large, protection systems are needed. Side effects from the machine protection systems compromising operational efficiency must be minimized.

\begin{figure}
\centering
\includegraphics[width=0.75\linewidth]{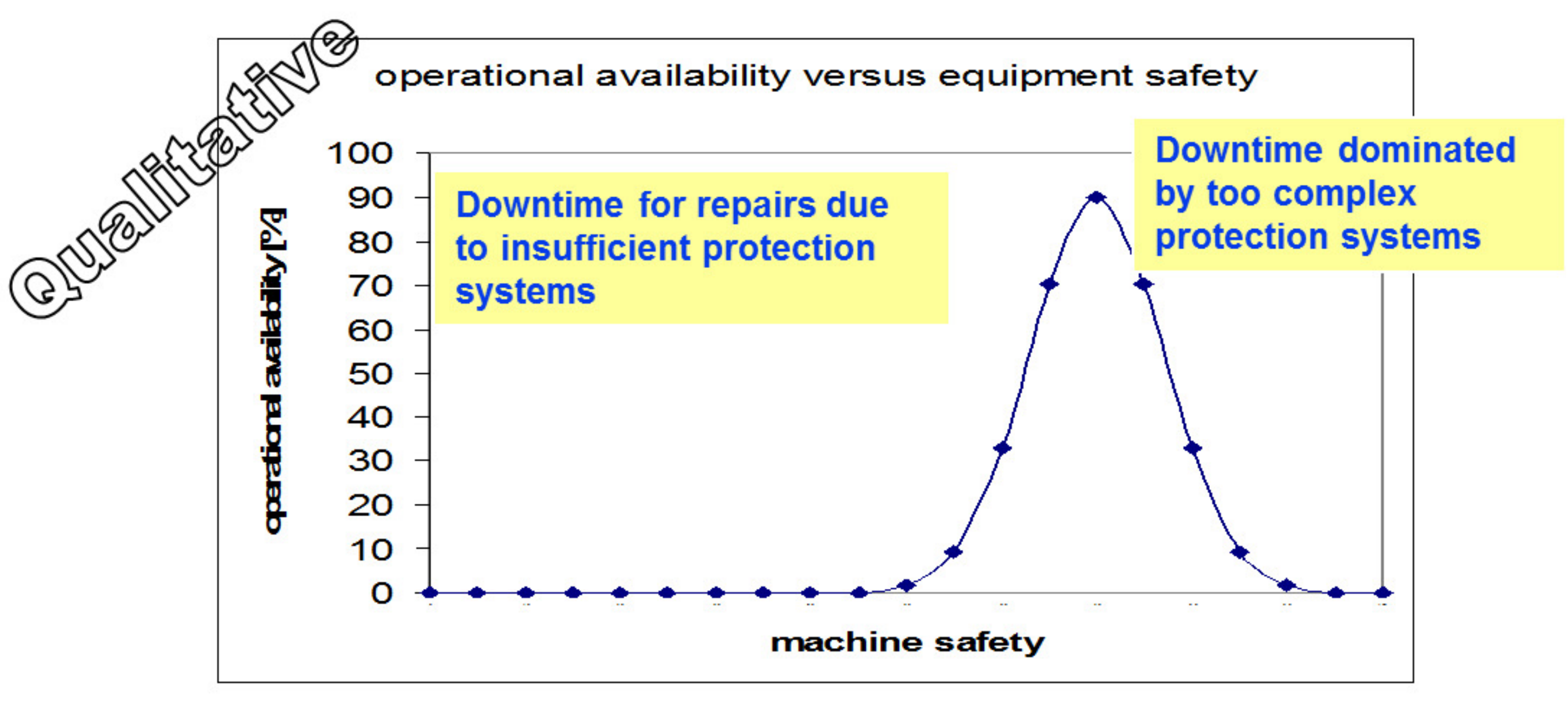}
\caption{Availability versus protection (qualitative graph). For too little protection, failures will lead to damage and reduce availability. For too much protection, the accelerator risks not being able to switch on and the availability is zero. There is some optimum in between.}
\label{Availability-versus-safety}
\end{figure}

Figure \ref{Availability-versus-safety} illustrates the availability for operation as a function of machine protection, for an accelerator with high beam power/large stored beam energy and considerable risk. If there is no protection system, failures will lead to damage and reduce availability. If there are too many and too complex protection systems, the accelerator risks not being able to switch on due to too many interlocks.

\section{Hazards and risks}

A hazard is a situation that poses a level of threat to the accelerator. Hazards are dormant or potential, with only a theoretical risk of damage. Once a hazard becomes `active' it becomes an incident or accident. An accident is defined as an unfortunate incident that happens unexpectedly and unintentionally, typically resulting in damage or injury.

Risk is a quantity that allows us to measure the threat of a hazard, by multiplying consequences and probability for a hazard becoming active:

\begin{equation}
    {\rm Risk} = {\rm Consequences} \times {\rm Probability}.
\end{equation}

Related to accelerators, consequences and probability of an uncontrolled beam loss need to be estimated to evaluate the risk. Machine protection systems prevent damage to equipment and reduce the risk, either by preventing a failure from occurring, or by mitigating the consequences of a failure. The higher the risk, the more effort is needed to prove that the protection system is sufficiently robust. Machine protection needs to be considered during design, construction and operation of the accelerator.

In the following sections, we discuss different types of hazards:

\begin{itemize}
\item Particle beams and their effects;
\item Electromagnetic energy stored in magnets and RF systems;
\item Other sources of energy.
\end{itemize}

\subsection{Hazards related to accelerator systems}

Even without operating with beam there can be hazards to be considered.

\begin{itemize}
\item A large amount of energy is stored in superconducting magnets. A complex magnet protection system is required \cite{Pfeffer2014}.
\item Powering normal-conducting magnets can overheat the magnets if water cooling fails. Monitoring of the temperature is required and interlocks to verify if the cooling works correctly. In case of a cooling problem the power converter is switched off \cite{Pfeffer2014}.
\item Power in RF systems (modulators, klystrons, wave-guides, cavities):  the presence of high-voltage arcs can damage the structure. Protection requires complex and fast interlock systems. For high-intensity beams, the RF system has to cope with fast transitions from beam on to beam off \cite{Kim2014}.
\item For high-voltage systems (e.g. kicker magnets) there is always the risk of arcing.
\item Powering systems (power converters, power distribution, electrical network \cite{Pfeffer2014}).
\end{itemize}

\subsection{Hazards related to particle beams}

\begin{figure}
\centering
\includegraphics[width=0.8\linewidth]{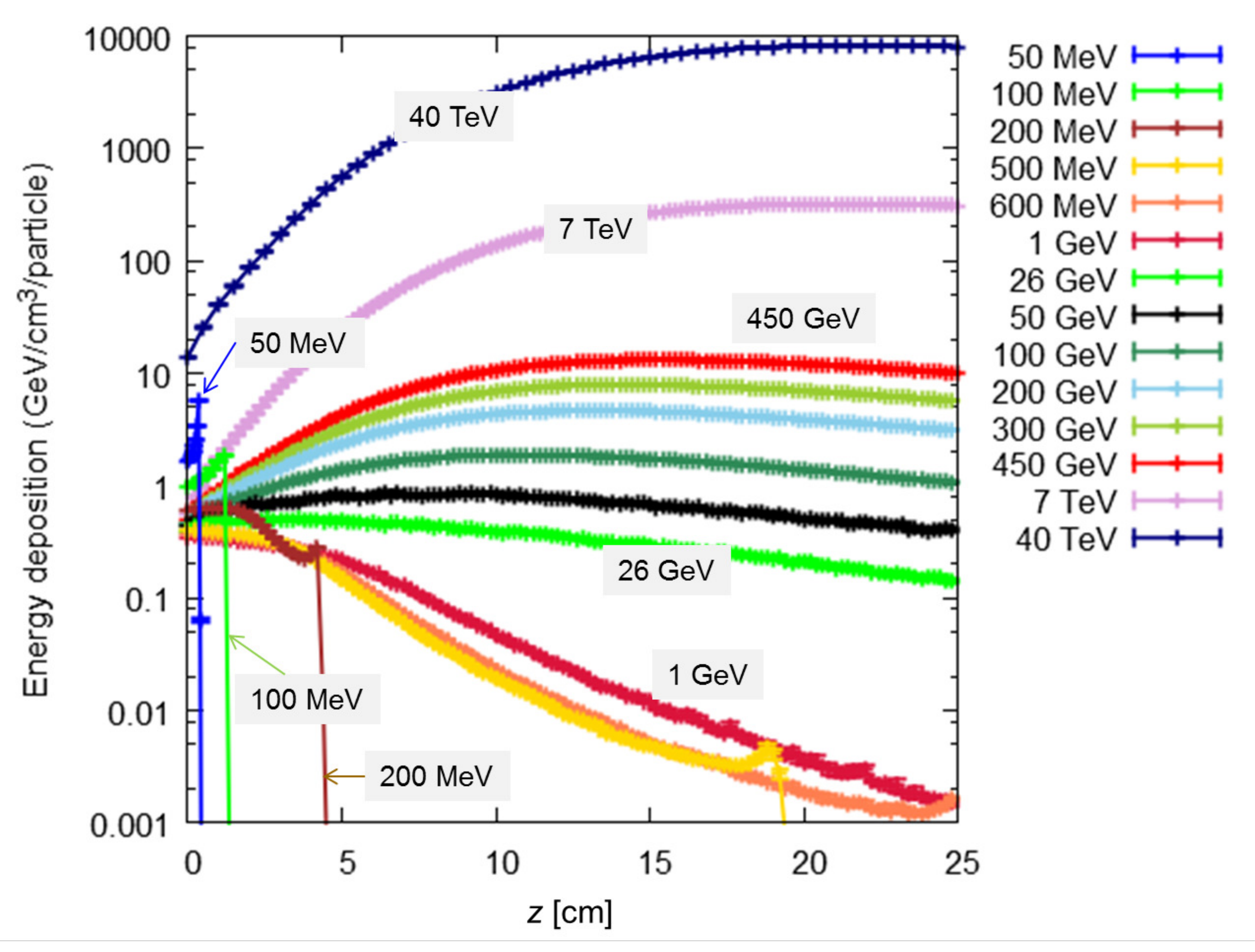}
\caption{Proton energy deposition for different energies for the impact of protons on copper \cite{Burkart2015}}
\label{Proton-Energy-deposition}
\end{figure}

When high-energy particles traverse matter, the particles deposit energy. The energy deposition depends on material, particle type and momentum. In \Fref{Proton-Energy-deposition}, the energy deposition of one proton as a function of depth is shown for a wide range of energies \cite{Burkart2015}. The energy deposition leads to a temperature increase. Material can change its material properties, deform, melt, or vaporize. Equipment can become activated. Superconducting magnets can quench (a quench is the transition from superconducting state to normal-conducting state).

\begin{itemize}
\item Regular beam losses during operation need be considered, since beam losses lead to activation of equipment and possibly quenches of superconducting magnets. Beam losses can also cause radiation-induced effects in electronics (single-event effects).
\item Accidental beam losses due to failures: understand hazards, e.g. mechanisms for accidental beam losses. Hazards become accidents due to a failure; machine protection systems mitigate the consequences.
\item Understand effects from synchrotron radiation that potentially lead to damage of equipment.
\item It is required to understand the interaction of particle beams with the environment, due to the impedance of the vacuum chamber and other accelerator equipment around the beam (kicker magnets, cavities, collimators, etc). This can lead to heating of equipment close to the beam and possibly to damage.
\end{itemize}

Even if the beam power or stored energy is not large, small beam sizes can cause localized damage. If the beam is lost, say, in a superconducting cavity, the consequences can be serious, as SNS (Spallation Neutron Source at Oak Ridge Laboratory) demonstrated \cite {Plum2014}.

Particle energy and particle type play an important role. For hadron accelerators beam losses lead to activation of the accelerator equipment. The emitted power of synchrotron radiation is very small and does not lead to a hazard. For electrons and positrons, activation is small, but the emitted power can be very high.

Not only the intensity, but also the beam/bunch structure (number of bunches, repetition frequency and bunch length) determines the interaction with the surrounding equipment due to the interplay of the electromagnetic fields with these structures.

\section{Strategy for machine protection}

\begin{figure}
\centering
\includegraphics[width=0.7\linewidth]{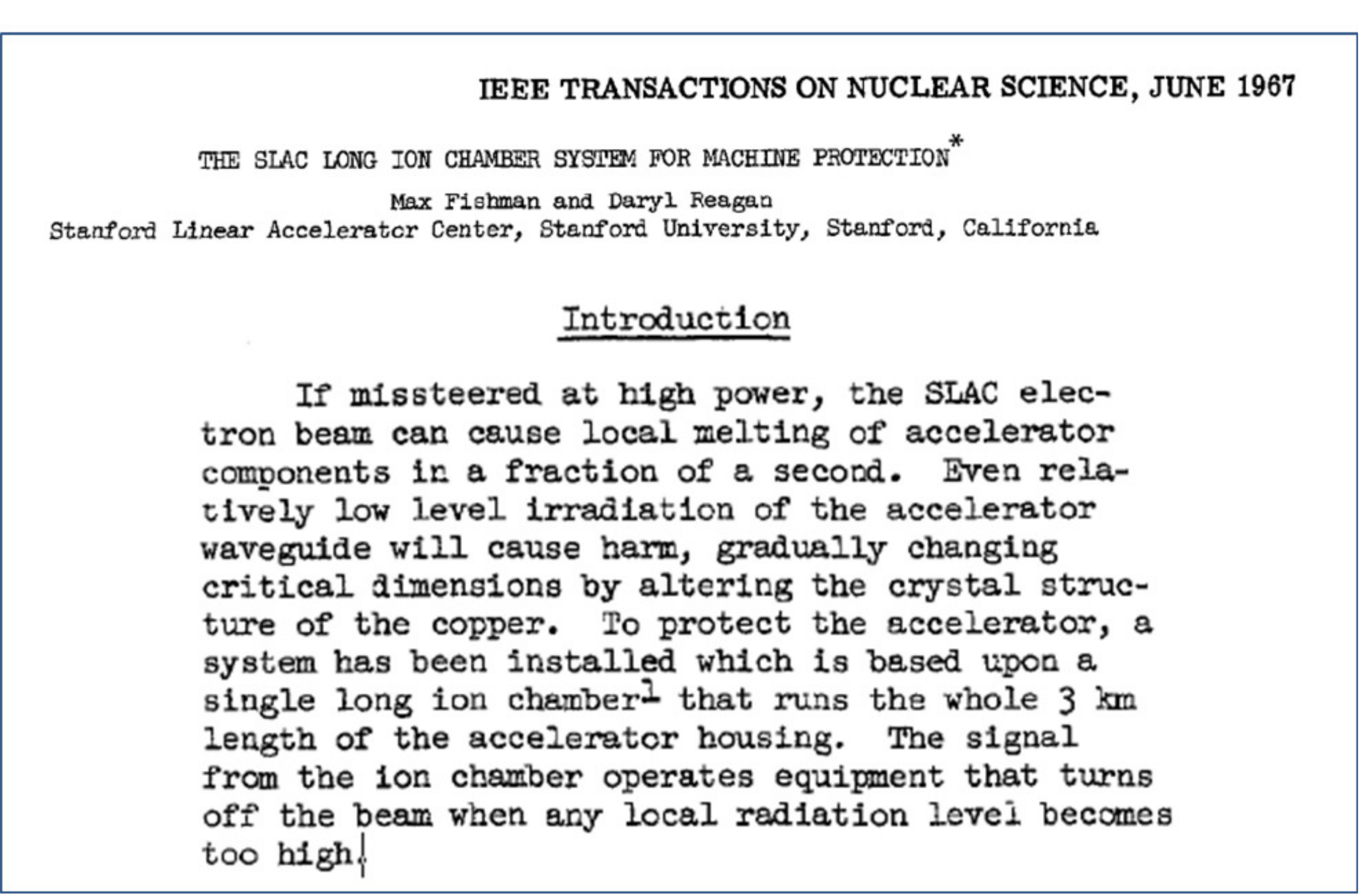}
\caption{Title and abstract of a very early paper on accelerator protection \cite{Fishman1967}}
\label{SLAC-first-paper}
\end{figure}

Machine protection has been on the agenda for a long time, but only in recent years has it become a significant topic at particle accelerator conferences. A very early paper was published in 1967: The SLAC long ion chamber system for machine protection \cite{Fishman1967}, already illustrating the risks when operating with high-power beams (\Fref{SLAC-first-paper}).

There are some principles for machine protection that need to be considered (sometimes this is referred to the $P^3$ rule for machine protection):

\begin{itemize}
\item Protect the machine;
\item Protect the beam;
\item Provide the evidence.
\end{itemize}

\noindent \textbf{Protect the machine:} The highest priority is to avoid damage to accelerator equipment.

\noindent \textbf{Protect the beam:} The objective is to maximize beam time, but complex protection systems reduce the availability of the machine. The number of `false' interlocks stopping operation must be minimized. This is a trade-off between protection and operation. A `false' interlock or `false' beam dump is defined as an interlock that stops operation even though there is no risk (example: a temperature sensor reading a wrong value, therefore switching off the power converter of a magnet and stopping beam operation).

\noindent \textbf{Provide the evidence:} If the protection systems stop operation (e.g. dump the beam or inhibit injection), clear diagnostics should be provided \cite{Zerlauth2009}. If something goes wrong (leading to damage, but also a near miss), it should be possible to understand the event. This needs synchronized transient recording of all the important parameters in all relevant systems, as well as long-term logging of parameters with reduced frequency (such as 1~Hz). Examples are the currents in all magnets, beam position, beam losses and beam intensity. The frequency of transient recording depends on the system and can be between Hz and MHz.

For beam operation, a list of all possible failures that could lead to beam loss into equipment should be considered. This is not obvious, since there is a nearly infinite number of mechanisms for losing the beam. However, the most likely failure modes and in particular the worst-case failures and their probabilities must be taken into account for the design of the protection system.

For a specific failure, the consequences of the failure need to be estimated, in terms of damage to equipment (repair requiring investment, e.g. in money), in downtime of the accelerator (e.g. in days) and in radiation dose to personnel accessing equipment (e.g. in mSv). In the estimation of downtime of the accelerator for repairs, the availability of spare parts needs to be considered. If the accelerator was operating for a long time with high-intensity beams, radioactive activation of material must be taken into account. It may be necessary to wait for cool-down of irradiated components to reduce the dose before accessing the equipment.

The second factor entering into the risk is the probability of such a failure happening (e.g. measured in number of failures per year).

\subsection{Active and passive protection}

In case of a failure, it takes a certain time until the beam is affected. The time constant is essential for the design of machine protection systems. We distinguish between two types of failures: failures where there is enough time to detect and mitigate (active protection), and failures where the time is too short for any mitigation (passive protection).

\noindent \textbf{Active protection} requires the detection of the failure by a sensor in an equipment system as early as possible, or by beam instrumentation detecting when the beam starts to be affected by the failure (for example, increased beam losses or a different trajectory). When a failure is detected, beam operation must be stopped. For synchrotrons and storage rings the beam is extracted by a fast kicker magnet into a beam dump block. The target must be designed to accept the beam pulse without being damaged. Injection must be stopped. For linacs, the beam is stopped in the low-energy part of the accelerator by switching off the source, by deflecting the low-energy beam by electrostatic plates (`choppers'), or by switching off the Radio Frequency Quadrupole (RFQ) for proton linacs. In the case of an accelerator complex with a chain of several accelerators, injection of beam into the next stage of the accelerator complex should be prevented.

Experience from LHC shows that for most types of failures a careful and fast monitoring of hardware parameters allows stopping beam operation before the beam is affected. Parameters monitored include state signals, other parameters, etc. As an example, a trip of a magnet power converter should be detected as early as possible.

It is not always possible to detect failures at the hardware level. The second method is to detect the initial consequences of a failure with beam instrumentation and to stop the beam before equipment is damaged. This requires reliable beam instrumentation. An electronic system (beam interlock system) links the different protection systems. It ensures that the beam is extracted from a synchrotron, injection is stopped, or RF acceleration might be stopped (for linacs). The interlock system might include complex logic that depends on the operational state.

\noindent \textbf{Passive protection:} There are failures (e.g. ultra-fast losses) when active protection is not possible. One example is the protection against mis-firing of an injection or extraction kicker magnet. A beam absorber or collimator is required to stop the mis-kicked beam in order to avoid damage. All possible beam trajectories for such failures must be considered and the absorbers must be designed to absorb the beam energy without being damaged. Another example is a fast extraction of a high-intensity beam from a circular accelerator into a transfer line. When the extraction takes place, the parameters of the transfer line, e.g. the current of the magnets, must be correctly set since for a wrong magnet current the beam would be deflected and possibly damage the vacuum chamber and other components.

There are a certain number of failures that can be completely eliminated. As an example, fast diagnostic kicker magnets that could deflect the beams into the vacuum chamber wall should only be installed in high-intensity machines if they are indispensable.

\section{Accidents at accelerators: looking into the past}

In this section and other lectures in this school, some examples of accidents are given:

\begin{itemize}
\item Damage to silicon detector in UA2 at the SPS proton--antiproton collider \cite{Beuville1989};
\item Vacuum chamber in SPS extraction line incident (see presentation in this school \cite{Kain2014});
\item Tevatron proton--antiproton collider (see presentation in this school \cite{Mokhov2014});
\item LHC magnet powering (see presentation in this school \cite{Wenninger2014});
\item CERN-LINAC 4 during commissioning at 3 MeV LINAC 4 (2013) at very low energy: beam hit a bellow and a vacuum leak developed;
\item J-PARC radioactive material leak accident (see presentation in this school \cite{Rokni2014}).
\end{itemize}

\begin{figure}
  \centering
  \includegraphics[width=0.8\linewidth]{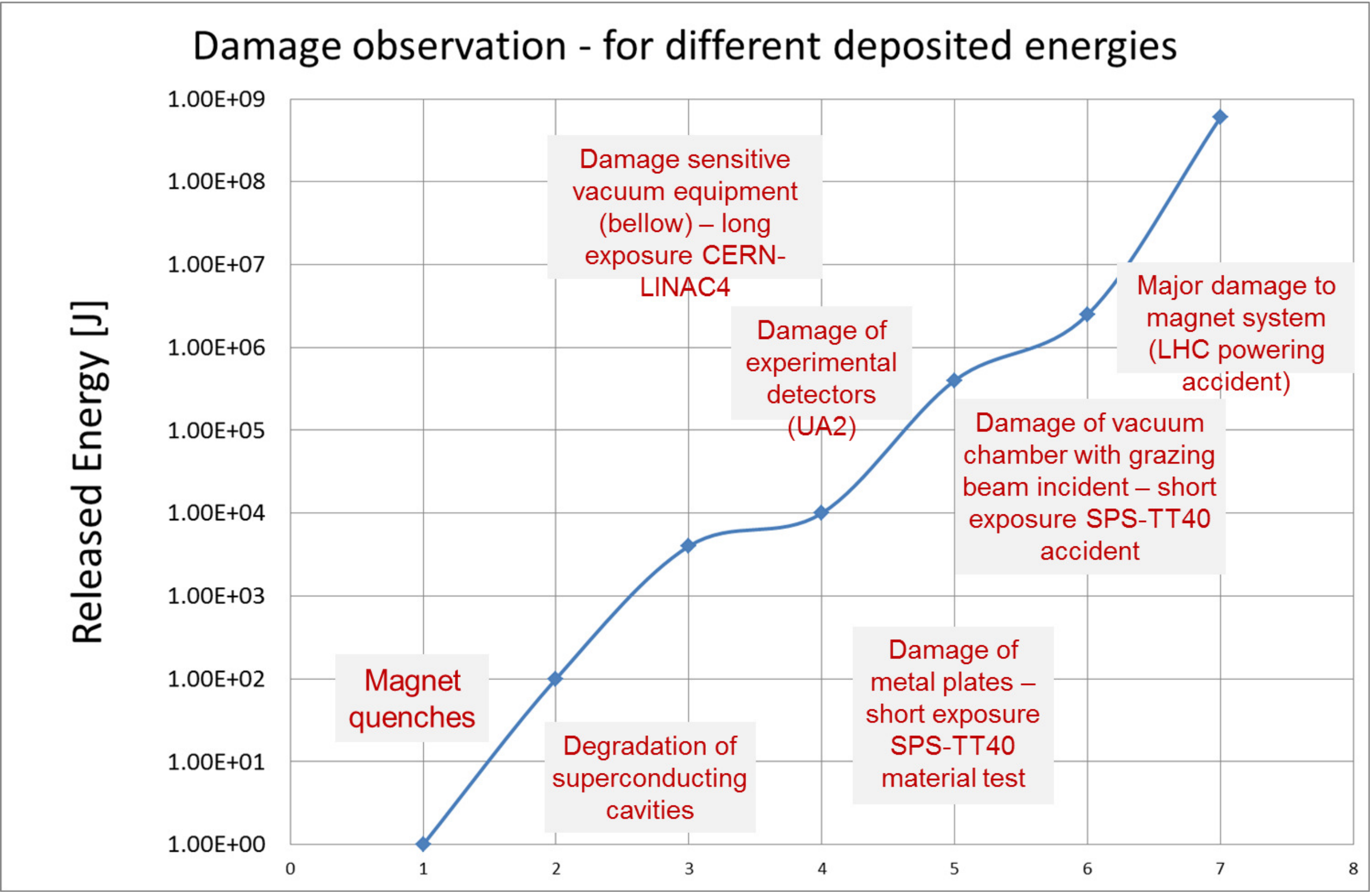}
  \caption{Deposited energy during quenches of magnets and damage of components for different events}
  \label{damage-summary}
\end{figure}

In \Fref{damage-summary}, the deposited beam or magnetic energy is shown for different events. It ranges from about 1~J sufficient to quench superconducting magnets to several 100~MJ that caused major damage to the LHC for the accident during powering tests in 2008.

\subsection{Damage to silicon detector in UA2 at the SPS proton--antiproton collider}

The silicon detector in the UA2 experiment at the SPS proton--antiproton collider was damaged during the injection process. Because of beam--beam effects, the beams needed to be separated at injection energy of 26 GeV and during the energy ramp to 315 GeV. This was done with electrostatic separators; their strengths were also ramped with the energy.

\begin{figure}
  \centering
  \includegraphics[width=0.8\linewidth]{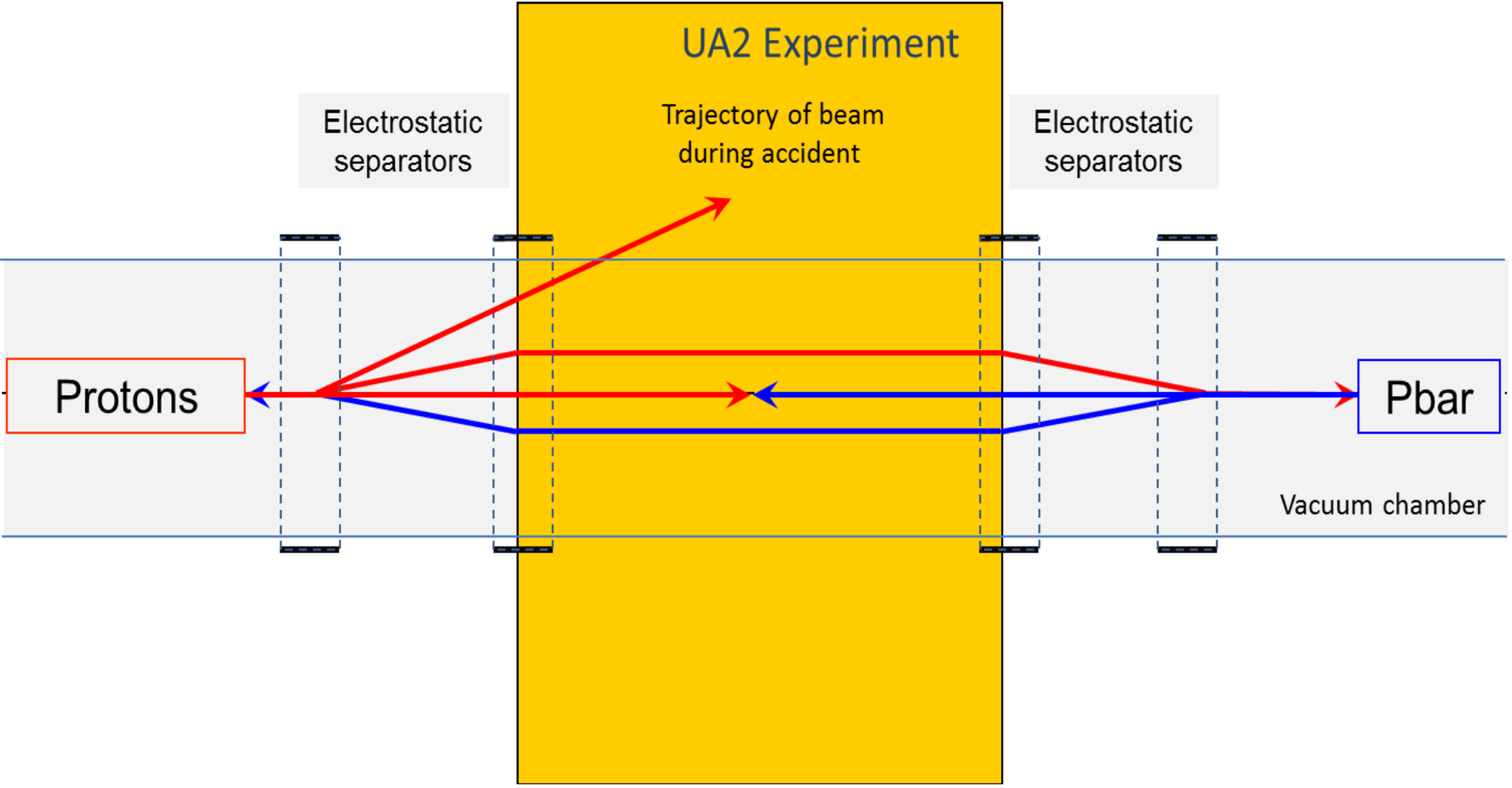}
  \caption{Illustration of trajectories of proton and antiproton beams for collisions, during injection and during the accident.}
  \label{UA2-accident}
\end{figure}

One day, during injection at 26~GeV, the separators were left accidentally at the settings for 315~GeV corresponding to a much too large angle and the beam was injected but not circulating, see \Fref{UA2-accident}. The separators directed the beam directly into UA2 using part of the detector as a beam dump during some minutes until the problem was understood \cite{Beuville1989}.

\subsection{CERN-LINAC4 during commissioning at 3 MeV}

\begin{figure}
  \centering
  \includegraphics[width=0.75\linewidth]{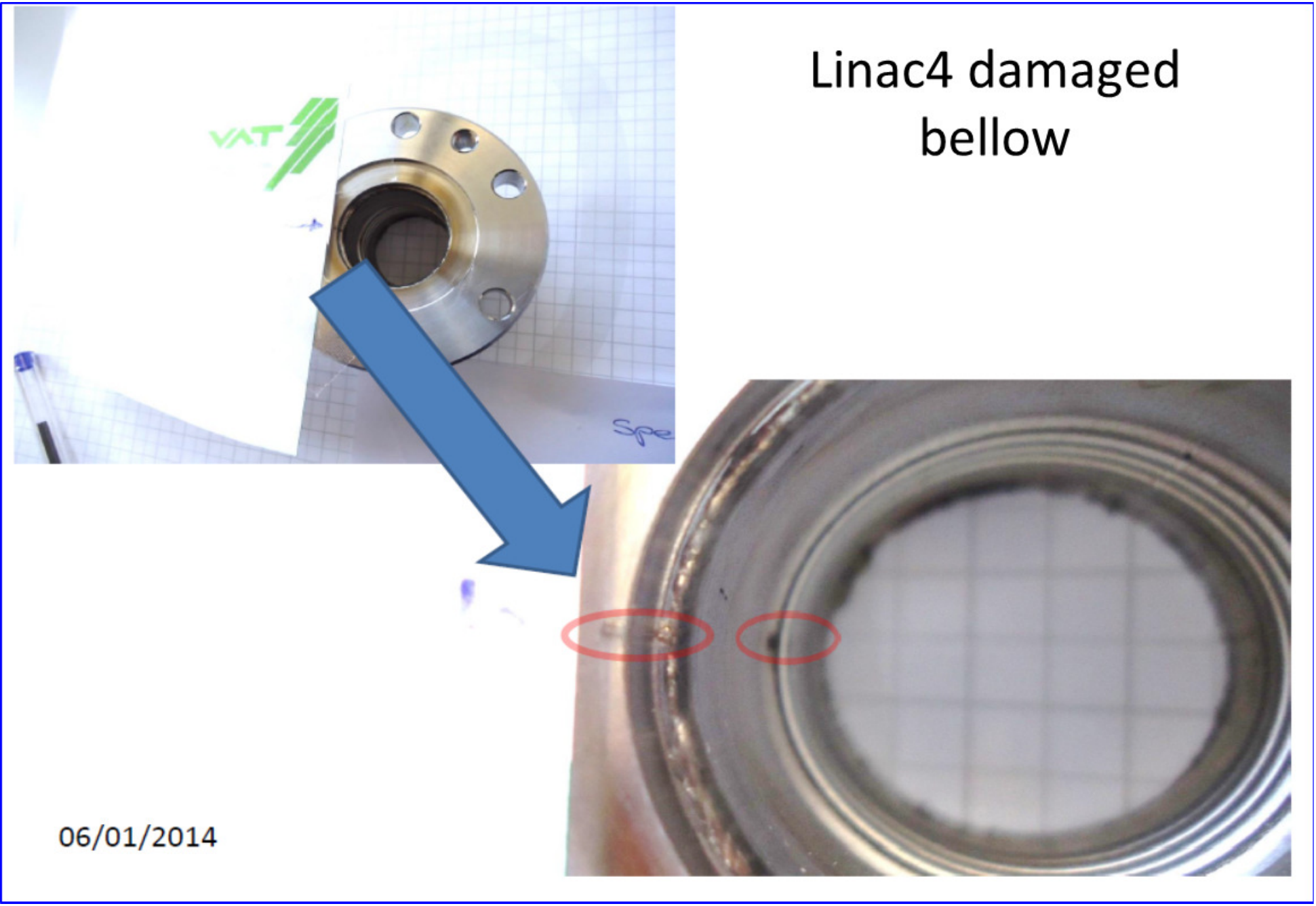}
  \caption{Damaged bellow showing signs of damage. After the damage a vacuum leak was observed and repair was required.}
  \label{LINAC4-accident}
\end{figure}

On 12 December 2013 a vacuum leak on a bellow developed in the MEBT (medium energy beam transfer) line (see \Fref{LINAC4-accident}). The analysis showed that the beam has been hitting the bellow during a special measurement with very small beams in the vertical plane but large in the horizontal plane. About 16\% of the beam was lost for about 14 min and damaged the bellow. The consequences were minor since LINAC4 is still being commissioned and not used in the chain of LHC injectors. The event demonstrates that even beams with very low power can cause damage.

\subsection{LHC magnet powering accident in 2008}

During the first phase of operation between 2009 and 2013 the magnetic field in the dipole magnets was limited, and therefore LHC was operating with a momentum of up to 4~TeV/$c$ and the maximum stored beam energy was up to about 140~MJ. This was the consequence of the 2008 LHC accident that happened during test runs without beam. A magnet interconnect was defective and the circuit opened. An electrical arc provoked a helium pressure wave damaging about 600~m of the LHC and polluting the beam vacuum over more than 2~km. An overpressure from the expansion of liquid helium damaged the structure. A total of 53 magnets had to be repaired. A detailed description of the accident is given in \cite{Wenninger2014}.

\section*{Acknowledgements}

I wish to thank many colleagues from CERN, ESS and the authors of the listed papers for their help and for providing material for this paper.

%\section{Bibliography}
%\bibliographystyle{ieeetr}
%\bibliography{\myreferences/Rudi-bibliography,\myreferences/Tahir,\myreferences/SPS-bibliography,\myreferences/Other-papers}

\end{document}